\documentclass[aps,prb,superscriptaddress,twocolumn,floatfix]{revtex4-1}

\usepackage{graphicx,color}
\usepackage{dcolumn}
\usepackage{bm}
\usepackage{stmaryrd}
\usepackage{latexsym}
\usepackage{amssymb}
\usepackage{amsfonts}
\usepackage{amsmath}

\usepackage{subfigure}

\usepackage{verbatim}

\begin{document}

\title{Electronic and spin dynamics in the insulating iron pnictide NaFe$_{0.5}$Cu$_{0.5}$As}

\author{Shunhong Zhang}
\affiliation{Institute for Advanced Study, Tsinghua University, Beijing 100084, China}
\affiliation{Department of Materials Science and Engineering, University of Utah, Salt Lake City, UT 84112, USA}

\author{Yanjun He}
\affiliation{Institute for Advanced Study, Tsinghua University, Beijing 100084, China}

\author{Jia-Wei Mei}
\email{meijw@sustc.edu.cn}
\affiliation{Institute for Quantum Science and Engineering, and Department of Physics, Southern University of Science and Technology, Shenzhen 518055}
\affiliation{Department of Materials Science and Engineering, University of Utah, Salt Lake City, UT 84112, USA}

\author{Feng Liu}
\email{fliu@eng.utah.edu}
\affiliation{Department of Materials Science and Engineering, University of Utah, Salt Lake City, UT 84112, USA}
\affiliation{Collaborative Innovation Center of Quantum Matter, Beijing 100084, China}

\author{Zheng Liu}
\email{zheng-liu@tsinghua.edu.cn}
\affiliation{Institute for Advanced Study, Tsinghua University, Beijing 100084, China}
\affiliation{Collaborative Innovation Center of Quantum Matter, Beijing 100084, China}

\date{\today}


\begin{abstract}
NaFe$_{0.5}$Cu$_{0.5}$As represents a rare exception in the metallic iron pnictide family, in which a small insulating gap is opened. Based on first-principles study, we provide a comprehensive theoretical characterization of this insulating compound. The Fe$^{3+}$ spin degree of freedom is quantified as a quasi-1D $S=\frac{5}{2}$ Heisenberg model. The itinerant As hole state is downfolded to a $p_{xy}$-orbital hopping model on a square lattice.  A unique orbital-dependent Hund's coupling between the spin and the hole is revealed. Several important material properties are analyzed, including (a) factors affecting the small $p-d$ charge-transfer gap; (b) role of the extra interchain Fe; and (c) the quasi-1D spin excitation in the Fe chains. The experimental manifestations of these properties are discussed.
\end{abstract}

\maketitle

\section{Introduction}
While the physics of high-temperature cuprate superconductors is generally attributed to doping a Mott insulator, \cite{RMP2006} the origin of iron-based superconductivity appears barely related. \cite{wang2012ironSC} Interestingly, it was recently found that by Cu substitution the iron pnictide superconductor NaFe$_{1-x}$Cu$_x$As exhibits Mott-insulating-like behavior, \cite{{Chen2013XAS},{Wang2015STM}} which provides a rare example bridging these two intriguing classes of superconductors. Indeed, scanning tunneling spectroscopy revealed striking similarities between the local electronic structure of NaFe$_{1-x}$Cu$_x$As and lightly doped cuprates. \cite{Wang2015STM} More recently, the $x$=0.5 limit, i.e. NaFe$_{0.5}$Cu$_{0.5}$As, was reached, in which Cu atoms was found to form well-ordered nonmagnetic 1D chains while the Fe atoms form 1D antiferromagnetic (AFM) chains. \cite{{song2016},{PRL2016ARPES}} Such a stoichiometric insulating sample largely excludes an insulating phase orignated from the Anderson localization. Angle-resolved photoemission spectroscopy (ARPES) revealed a narrow band gap of the size $\sim$16 meV, which was further examined by density functional theory (DFT) calculation plus the onsite U correction (DFT+U). \cite{PRL2016ARPES}

Considering that the gap size is comparable to that in a narrow-gap semiconductor, charge excitations are expected to remain active at ambient temperature. In addition, the magnetically ordered quasi-1D Fe chains should support unique spin excitations, which might provide clues to understand the interplay between AFM magnetic order and superconductivity in Fe-based superconductors. \cite{RMP2015Dai} This article aims to provide a systematic description of the low-energy physics in NaFe$_{0.5}$Cu$_{0.5}$As within the DFT+U formalism. The paper is organized as follows. Section II describes the methodology. Section III reproduces the DFT+U results based on the experimentally determined chain-like structure, and further clarifies the charge-transfer nature of the energy gap and the spin state of each element. Section IV studies how the electronic structure changes when this chain structure is perturbed. This result indicates a close connection between the insulating phase and the formation of quasi-1D AFM chains. It also explains the robustness of this insulating phase when iron concentration increases. Section V quantifies the effective spin model and discusses its manifestation in experiment. In Section VI, we reveal unique orbital-dependent spin polarization of the hole bands due to its coupling to the AFM Fe chains. Section VII concludes this article.

\begin{figure*}
\centering
\subfigure{
\begin{minipage}{0.25\textwidth}
\includegraphics[height=8.4cm]{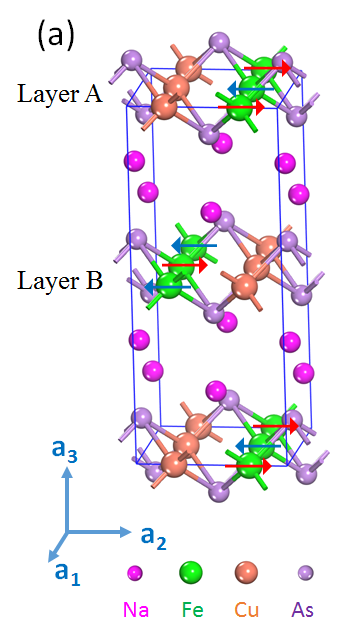}
\label{fig-1a}
\end{minipage}}
\subfigure{
\begin{minipage}{0.3\textwidth}
\includegraphics[height=8cm]{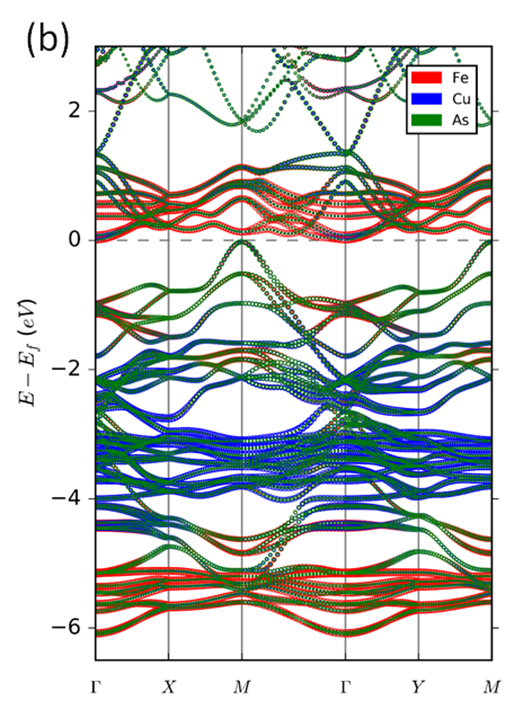}
\label{fig-1b}
\end{minipage}}
\hspace{0.2cm}
\subfigure{
\begin{minipage}{0.3\textwidth}
\includegraphics[height=3cm]{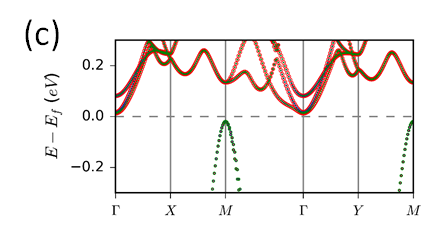}\\
\label{fig-1c}
\includegraphics[height=5cm]{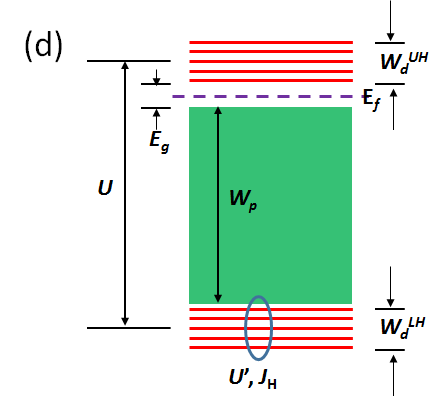}
\end {minipage}}
\subfigure{\label{fig-1d}}
\caption{(a) Crystal and magnetic structure of NaFe$_{0.5}$Cu$_{0.5}$As. Red arrows denote the spin direction. (b) Electronic band structure of NaFe$_{0.5}$Cu$_{0.5}$As, with the atomic composition projection. (c) Zoom in plot of (b) near the Fermi level. (d) Schematic energy diagram. The key energy scales are summarized in Table \ref{table-1}.}
\label{fig-1}
\end{figure*}

\section{Calculation method}

The experimentally determined structure of NaFe$_{0.5}$Cu$_{0.5}$As contains alternatively aligned AFM Fe and nonmagnetic Cu chains along the [100] direction [Fig.\ref{fig-1a}], as revealed by the high resolution TEM measurements and neutron scattering \cite{song2016}. Starting from this lattice and magnetic structure, DFT+U calculations are performed using Vienna $Ab\ initio$ Simulation Package (VASP) \cite{Kresse1994}. The +U correction follows the simplified (rotational invariant) approach introduced by Dudarev \cite{Dudarev1998LDA+U}:
\begin{eqnarray}
\label {eq-1}
E_{DFT+U}=E_{DFT}+\frac{U_{eff}}{2}\sum_{\sigma,m}[n_{m,m}^\sigma-(\hat{n}^\sigma \hat{n}^\sigma)_{m,m} ],
\end{eqnarray}
where $m$ is the magnetic quantum number of the five Fe 3$d$-orbitals (for the present case), and $\hat{n}$ is the onsite occupancy matrix. This +U correction can be understood as adding a penalty functional to the DFT total energy expression that forces the $d$-orbitals either fully occupied or fully empty, i.e.,  $ \hat n^{\sigma} = \hat n^{\sigma} \hat n^{\sigma}$.  We set U$_{eff}=2.8$ eV, as used in the previous study to get the correct insulating gap size \cite{PRL2016ARPES}.

With respect to the DFT functional, electron exchange and correlation are treated by using the Perdew-Burke-Ernzerh functional \cite{GGA-PBE} with the projector augmented wave method. \cite{PAW} Plane wave basis sets with a kinetic energy cutoff of 300 eV is used to expand the valance electron wave functions. Monkhorst-Pack \cite{MP} k point grid of $8\times8\times4$ are adopted to represent the first Brillouin zone. The electronic and spin ground state is determined self-consistently until the energy threshold 10$^{-5}$ eV is reached.

\begin{figure*}
\centering
\subfigure{\label{fig-2a}
\includegraphics[height=8cm]{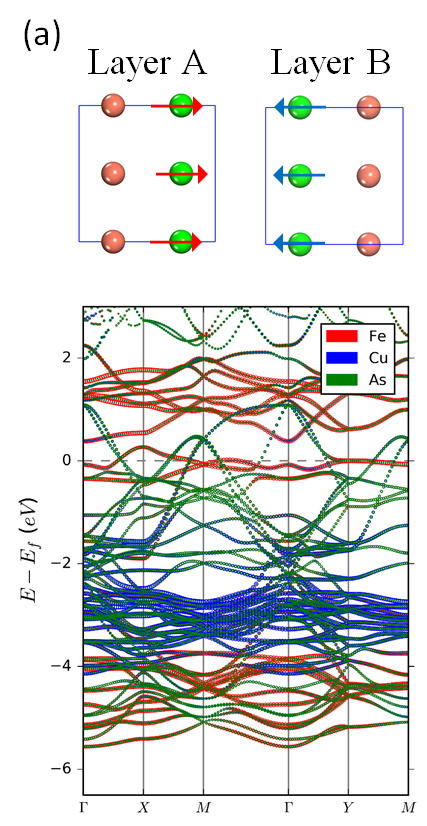}}
\subfigure{\label{fig-2b}
\includegraphics[height=8cm]{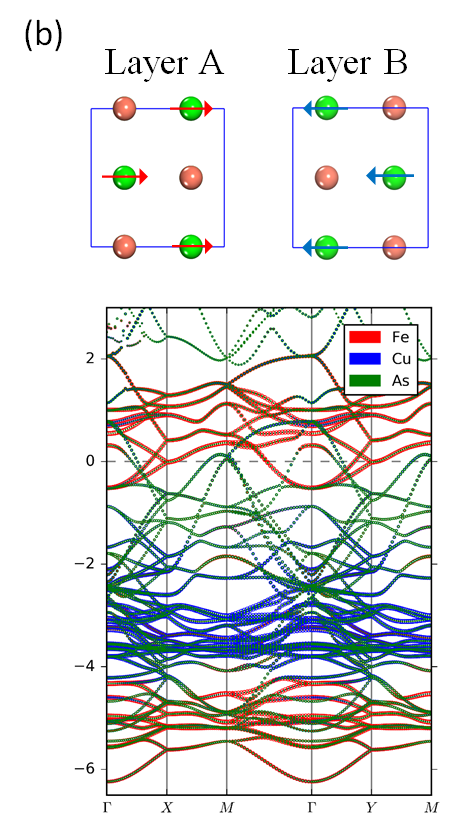}}
\subfigure{\label{fig-2c}
\includegraphics[height=8cm]{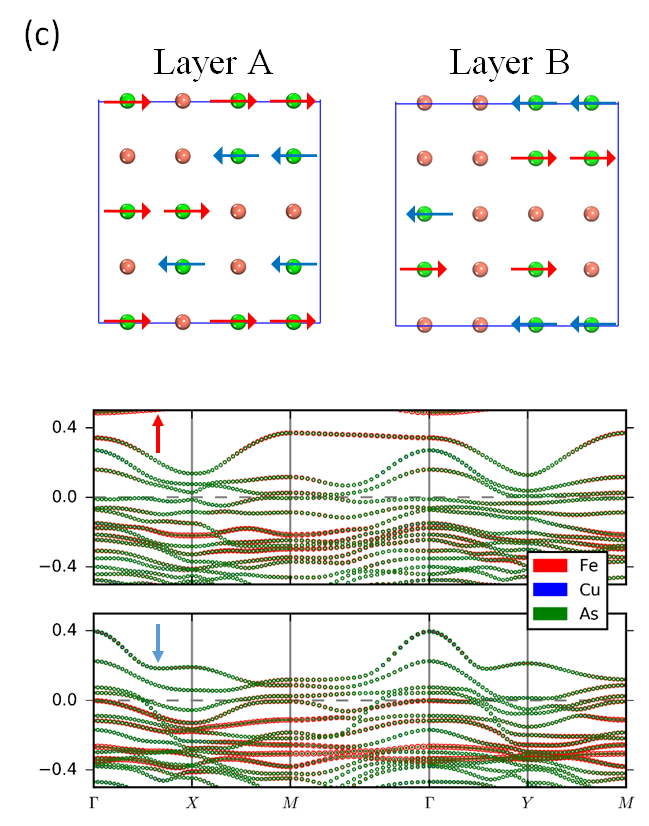}}
\caption{Three hypothetical atomic and magnetic structures and the corresponding band structures: (a) FM spin configuration ; (b) Fe and Cu ions form a checkerboard pattern; (c) Fe and Cu ions are assumed to distribute randomly. In all these cases, the charge-transfer gap is closed.}
\label{fig-2}
\end{figure*}

\section{The insulating ground state}

In Fig.\ref{fig-1a} and \ref{fig-1b} we plot the gound state atomic and magneitc structure (structural data from ref. \cite{song2016}) as well as the DFT+U band structure of NaFe$_{0.5}$Cu$_{0.5}$As, and mark each band with its chemical compositions. It is very different from the metallic iron pnictides as reflected by a small gap at the Fermi level [Fig.\ref{fig-1c}]. More importantly, the Fe bands split into two sets: one right above the Fermi level (upper Hubbard bands, UH), and the other deep inside the occupied states (lower Hubbard bands, LH). The occupied band edge consists of nearly pure As-orbitals free from strong correlation. This explains the surprisingly excellent agreement between DFT+U and ARPES results around the occupied band edge. \cite{PRL2016ARPES} The sharp difference of the chemical component between the occupied and unoccupied band edge also explains the strongly asymmetric dI/dV spectral line shape observed by STM when the bias reverses. \cite{Wang2015STM} We should point out that a previous DFT+U calculation attributed a large fraction of Cu-orbital contribution to the occupied band edge,\cite{PRL2016ARPES} whereas our result indicates that the majority of Cu bands stay deep below the Fermi level.

\begin{table}[ht]
\caption{Key energy scales extracted from the DFT+U calculation }
\label{table-1}
\begin{ruledtabular}
\begin{tabular}{cc}
Charge Sector - Fig. 1&  (eV) \\
  \hline
$W_p$  & 5.5 \\
$W_d$  & 1.0 \\
$E_g$ & 3.0$\times10^{-2}$ \\
$U$ & 6.0 \\
$U'$ & 3.5\\
$J_H$ & 1.2\\

\hline
Spin Sector - Fig. 4 & (meV) \\
\hline
$J_1$ & 30 \\
$J_2$ & -0.43 \\
$J_3$ & 0.14\\
\end{tabular}
\end{ruledtabular}
\end{table}

Both Fe- and Cu-dominated bands are narrow, distinct from the dispersive As bands, which reflects the localized nature of $d$-electrons. The valence states of Fe and Cu can then be determined by counting the number of occupied bands of each element. This analysis indicates a +1 valence state for Cu ($3d^{10}$) and a +3 valence state for Fe ($3d^5$), respectively, which is consistent with the experimental observation that only Fe atoms exhibit local magnetic moment. \cite{song2016} The half filled Fe $d$-bands all have the same spin polarization. Therefore, the Fe$^{3+}$ ion is in the high-spin state, effectively forming a S=5/2 moment. In comparison, in NaFeAs iron is in the +2 valence state ($d^{6}$). Therefore, Cu substitution of Fe can be effectively considered as hole doping. For NaFeAs,  the total spectral weight in the neutron scattering measurement suggests an effective S=1/2 local spin. \cite{PRL2014NaFeAs} The same measurement indicates a much larger local moment in NaFe$_{0.5}$Cu$_{0.5}$As, but still less than S=5/2 \cite{{song2016}}. There are many reasons that the experimentally determined value may differ from expected and the underlying reason is worth of further investigation.

A schematic energy diagram is drawn in Fig.\ref{fig-1d}. Several key energy scales can be readily extracted from Fig. \ref{fig-1b}, which are summarized in Table \ref{table-1}. The energy gap around the Fermi level ($E_g$) is of a $p-d$ charge transfer origin, just like in cuprates. \cite{RMP2006}. The green region indicates the itinerant As $p$-bands, which extend from around -6 eV up to the Fermi level, despite intertwining with the Fe LH bands and the Cu bands in between. The splitting between the UH and LH bands is determined by the intra-orbital Hubbard repulsion of Fe 3$d$-orbitals ($U$).

We note that the effective Coulomb repulsion $U_{eff}=2.8\ eV$ as set for the DFT+U calculation is defined as \cite{Dudarev1998LDA+U}:
\begin{eqnarray}
\label{eq-2}
U_{eff}&=&\overline{\langle mm' | V_{ee} | mm'\rangle} - \overline{\langle mm' | V_{ee} | m'm\rangle}_{m\neq m'}  \nonumber \\
&=&\frac{U+4U'}{5}-J_H ,
\end{eqnarray}
which is an average of the intra-orbital Coloumb repulsion ($U$ for $m=m'$) and inter-orbital repulsion ($U'$ for $m\neq m'$) minus the Hund's coupling $J_H$. It is possible to further determine $U'$ and $J_H$, by considering that $U$, $U'$ and $J_H$ are not independent. We assume that the screened Coulomb potential ($V_{ee}$) is still spherically symmetric, and it is known that the relation $U'+2J_H=U$ holds. \cite{PRB78U+J} Then, in combination with Eq. (\ref{eq-2}), the values of $U'$ and $J_H$ can be calculated (Table \ref{table-1}).

\begin{figure*}
\centering
\subfigure{
\begin{minipage}{0.32\textwidth}
\includegraphics[height=7cm]{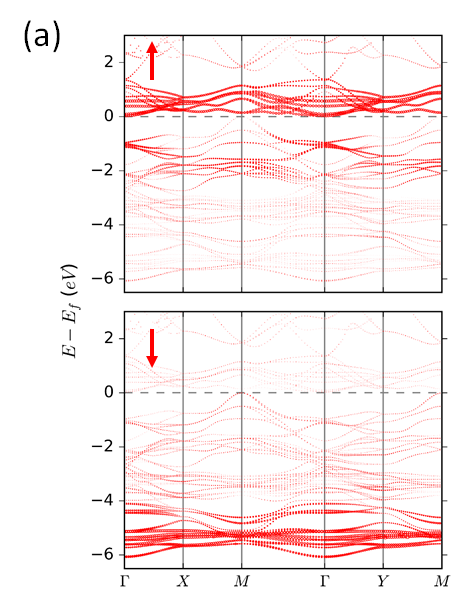}
\label{fig-3a}
\end{minipage}}
\hspace{0.1cm}
\subfigure{
\begin{minipage}{0.54\textwidth}
\includegraphics[height=7cm]{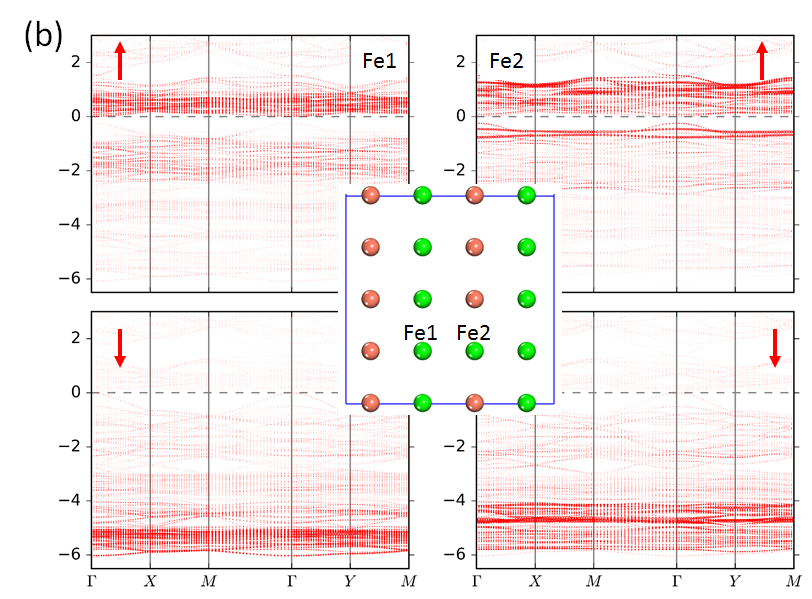}
\label{fig-3b}
\end{minipage}}
\caption{The effect of inter-chain Fe as shown by the band structure of (a) NaFe$_{0.5}$Cu$_{0.5}$As and (b) NaFe$_{0.53}$Cu$_{0.47}$As. The weight of the bands is proportional to the wavefunction projection on the Fe site.}
\label{fig-3}
\end{figure*}

\section{Factors affecting the charge-transfer gap}
The small charge-transfer gap arises from a delicate separation between the Fe UH band and the As $p$-band. Fig. \ref{fig-2a} shows that the gap is closed by enforcing a ferromagnetic spin configuration. We have also artificially rearranged the Fe/Cu atoms into a checkerboard pattern [Fig.\ref{fig-2b}] or randomly [Fig.\ref{fig-2c}]. In all the cases, the charge-transfer gap no longer exists. These results indicate the importance of the quasi-1D AFM chain structure to the observed insulating ground state. A recent DFT+dynamical mean-field theory calculation also found that the correct insulating ground state could not be reproduced without the quasi-1D AFM magnetic order. \cite{DMFT} Nevertheless, the splitting of the UH and LH Fe bands, which signifies the Mott localization of the $d$-electrons, is largely independent of the magnetic or atomic structure.

Another question is why this insulating phase appears much before the $x=0.5$ stoichiometric limit is reached. A recent work applying the real space Green's function method emphasized the role of disorder. \cite{PRB2015localization} Here, we would like to point out that the interchain Fe is in a different valence state. Within the DFT+U formalism, we have constructed a 2$\times$2$\times$1 supercell and replaced one of the Cu atoms with Fe. The DFT+U band structure indicates that the charge-transfer gap indeed remains open. Fig. \ref{fig-3b} show the orbital-resolved bands by projecting the Bloch wavefunctions onto the in-chain Fe (Fe1) and the inter-chain Fe (Fe2) ions. We can observe that Fe1 [Fig.\ref{fig-3b}, left] is half-filled as in NaFe$_{0.5}$Cu$_{0.5}$As [Fig.\ref{fig-3a}], whereas two additional occupied bands dominated by Fe2 can be found below the Fermi level [Fig. \ref{fig-3b}, right]. Based on this observation, the robustness of the gap to the extra Fe atoms can be explained as follows. The key point is that the in-chain Fe ($3d^5$) structure is not perturbed. These inter-chain Fe atoms are roughly in a ($3d^7$) state, which nominally loses one electron each, the same as the replaced Cu$^{1+}$ ion, and thus do not introduce extra charge carriers. We propose that the existence of two types of Fe ions can be verified by X-ray adsorption spectroscopy measurement.

\section{Spin exchange and excitation spectrum}
\begin{figure*}
\centering
\subfigure{\label{fig-4a}
\begin{minipage}{0.95\textwidth}
\includegraphics[height=6cm]{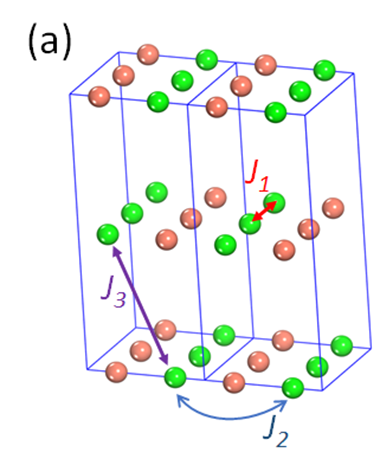}
\includegraphics[height=6cm]{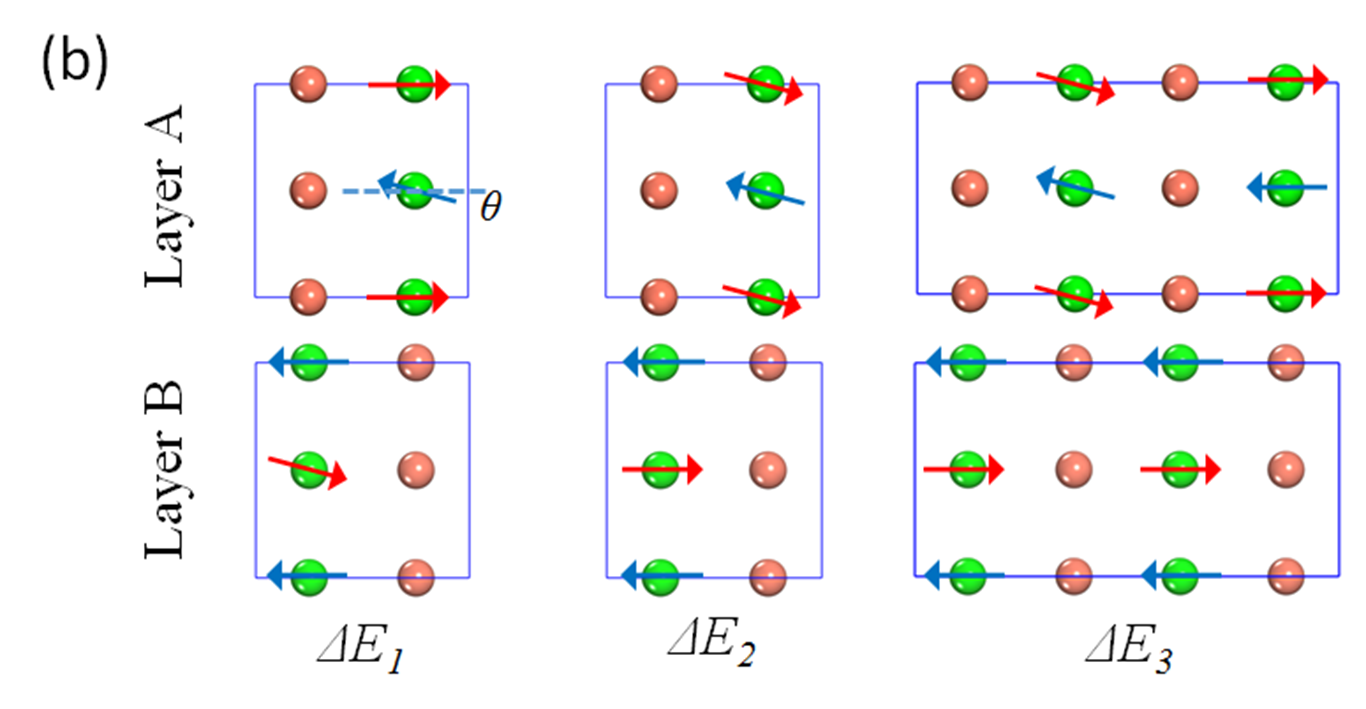}\\
\includegraphics[height=5.6cm]{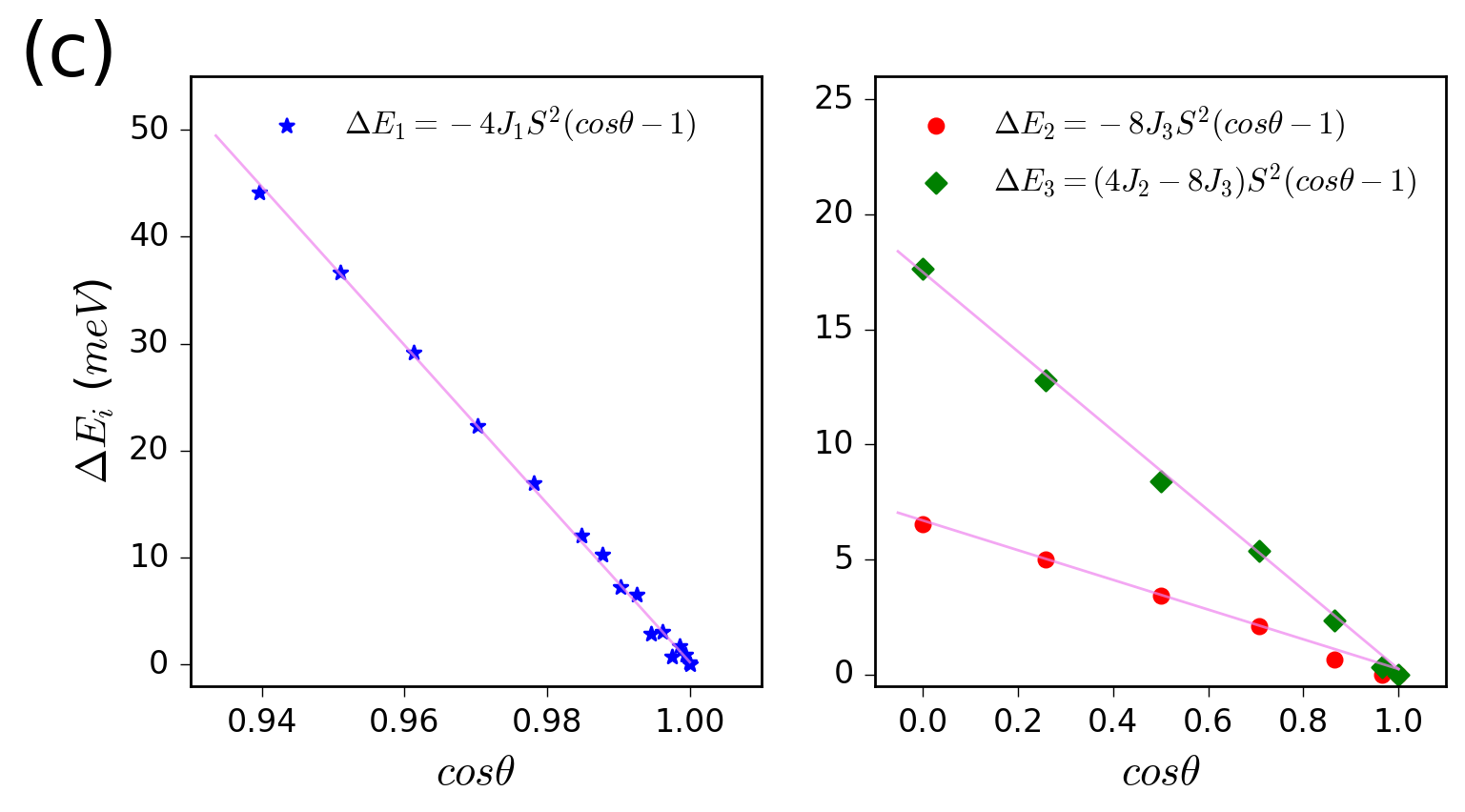}
\hspace{0.5cm}
\includegraphics[height=5.6cm]{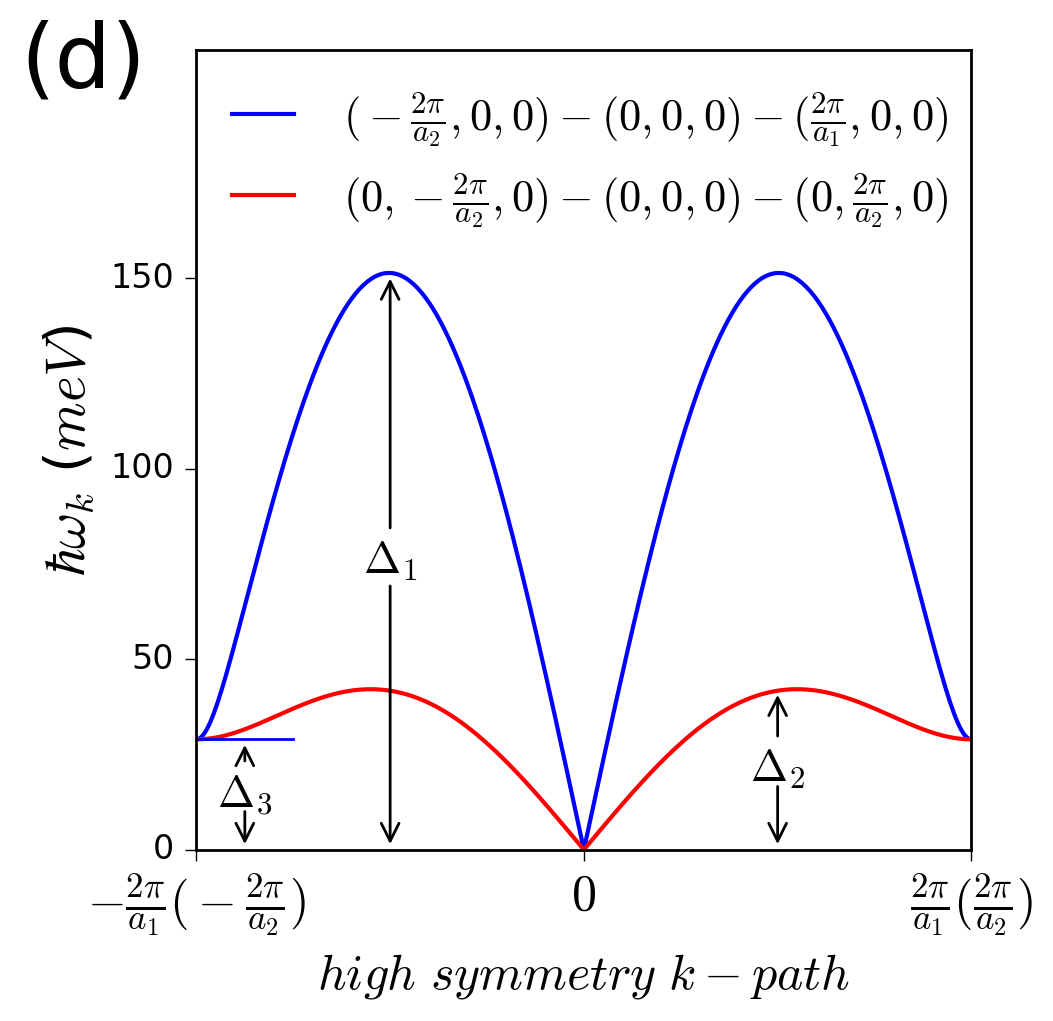}
\end{minipage}
}
\subfigure{\label{fig-4b}}
\subfigure{\label{fig-4c}}
\subfigure{\label{fig-4d}
\begin{minipage}{0.28\textwidth}
\end{minipage}
}
\caption{ (a) The primary nearest-neighbor (NN) exchange in Eq.(\ref{Hs}). (b) Spin rotation for extraction of $J_{1,2,3}$.  (c) Total energy as a function of the rotation angle. The fitted lines are also shown. (d) In-plane magnon dispersion.}
\label{fig-4}
\end{figure*}

After clarifying the Mott localization associated with the Fe$^{3+}$ ($3d^5$) electrons, the spin excitation in NaFe$_{0.5}$Cu$_{0.5}$As can be reasonably described by a S=$\frac{5}{2}$ spin model. We assume a Heisenberg-type model:
\begin{eqnarray} \label{Hs}
H_d=J_1\sum_i \textbf{S}_i \cdot \textbf{S}_{i+\frac{\textbf{a}_1}{2}}+J_2\sum_i \textbf{S}_i \cdot \textbf{S}_{i+{\textbf{a}_2}} \\ \nonumber
+J_3\sum_i (\textbf{S}_i \cdot \textbf{S}_{i+\frac{\textbf{a}_2}{2}+\frac{\textbf{a}_3}{2}}+\textbf{S}_i \cdot \textbf{S}_{i-\frac{\textbf{a}_2}{2}+\frac{\textbf{a}_3}{2}}),
\end{eqnarray}
where $\textbf{a}_{i=1,2,3}$ are the three lattice vectors as shown in Fig. \ref{fig-1a}. The in-chain exchange is expected to be the dominant spin-spin interaction. The inter-chain coupling is weak, yet important to forming the 3D magnetic order at finite temperature. For each dimension, we include the nearest-neighbor term only, as shown in Fig. \ref{fig-4a}.

The three exchange parameters $J_{1,2,3}$ are quantified by calculating the DFT+U total energy increase after applying a perturbation to the ground-state spin configuration. The choice of the perturbation should be as small as possible to avoid insulator-to-metal transition, but still numerically stable. To extract the in-chain exchange, we therefore rotate a single spin in each chain by a small angle $\theta$ [Fig. \ref{fig-4b}]. By adding a penalty term to the standard Kohn-Sham potential, our noncollinear spin-polarized DFT calculations are able to obtain the total energy of these excited magnetic configurations.\cite{constr_mag} We then assume a classical mapping between DFT+U total energy and Eq. (\ref{Hs}): $\Delta E_1(\theta)=-zJ_1 S^2(cos \theta-1)$, where $z$ is the number of perturbed bonds within the unit cell. Finally, $J_1$ is determined by a linear fitting between $\Delta E_1$ and $cos \theta$ [Fig.\ref{fig-4c}, left panel]. This method has been successfully implemented to study the spin excitation of other iron-based superconductors. \cite{xiang2016KFeSe} To extract the inter-chain exchange, we apply a global rotation to the spins in a single chain or a single Fe layer, and a similar linear fitting can be performed [Fig.\ref{fig-4b} and \ref{fig-4c}]. Due to the small magnitude of the inter-chain exchange, the rotation angle in these two cases should be much larger. Nevertheless, the perturbed states stay in the proximity of the magnetic ground state and the Mott physics does not change. For all the cases, the numerical data are found to be well reproduced by the linear fitting, which in turn justifies the Heisenberg-type exchange employed in Eq. (\ref{Hs}). We summarize the values of $J_{1,2,3}$ in Table \ref{table-1}. AFM exchange corresponds to a positive $J$, and FM exchange corresponds to a negative $J$.

The magnon spectrum $\omega(\textbf{k})$ can then be calculated by the standard spin-wave expansion \cite{Madelung2012}:
\begin{eqnarray}
\omega(\textbf{k})&=&\sqrt{A(\textbf{k})^2-B(\textbf{k})^2} \\ \nonumber
A(\textbf{k})&=&2SJ_1-2SJ_2[1-\cos({k_2a_2})]+4SJ_3 \\ \nonumber
B(\textbf{k})&=&2SJ_1\cos(\frac{k_1a_1}{2})+4SJ_3\cos(\frac{k_2a_2}{2})\cos(\frac{k_3a_3}{2})
\end{eqnarray}
Fig.\ref{fig-4d} shows the dispersion along in-plane high-symmetry directions in the momentum space. Along the chain, typical AFM spin wave can be observed. The band top is reached at $(\pi/a_1, 0, 0)$ with the energy $\Delta_1=S(2J_1+4J_3)\sim2SJ_1$. The magnon energy does not return to zero at ($\pm2\pi/a_1$,0,0) due to the out-of-plane exchange $J_3$. The energy gap is $\Delta_3=4S\sqrt{2J_1J_3}$. The weak interchain exchange mixing with $J_1$ also leads to noticeable dispersion perpendicular to the chain. Around $(0, \pm\pi/a_2, 0)$, the magnon energy is $\Delta_2\approx4S\sqrt{J_1(|J_2|+J_3)}$. According to the calculated values of $J_{1,2,3}$, $\Delta_2/\Delta_1=2\sqrt{\frac{|J_2|+J_3}{J_1}}\sim\frac{1}{4}$. Above $\Delta_2$, the constant energy contour as measured from inelastic neutron scattering experiment should display typical 1D features, in contrast to the low-energy anisotropic 2D topology. This energy scale reversely provides a way to determine the weak interchain exchange experimentally.

\begin{figure*}
\centering
\subfigure{\label{fig-5a}
\begin{minipage}{0.88\textwidth}
\includegraphics[height=5cm]{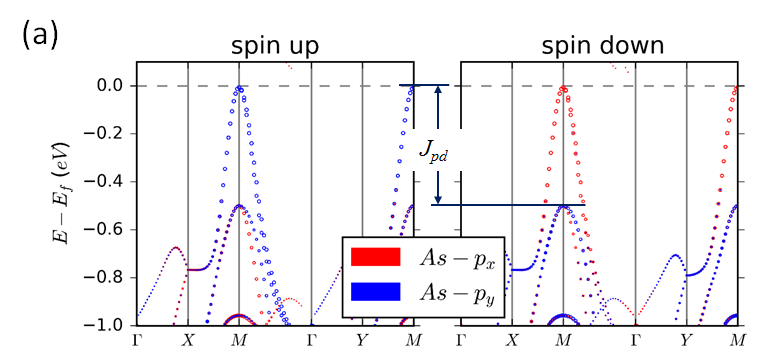}
\includegraphics[height=5cm]{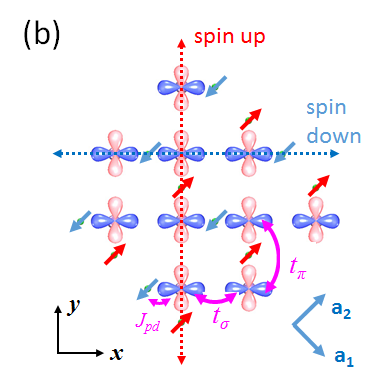}\\
\includegraphics[height=5cm]{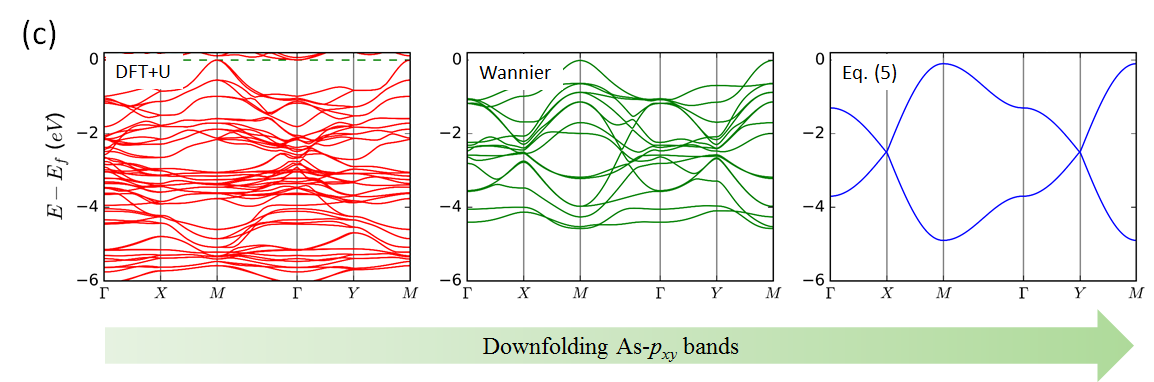}
\end{minipage}
}
\subfigure{\label{fig-5b}}
\subfigure{\label{fig-5c}}
\caption{Hoppings and spin polarization of the As-$p_{xy}$ orbitals and formation of the hole bands in NaFe$_{0.5}$Cu$_{0.5}$As. (a) Spin and orbital resolved valance bands. The bands are projected on the $p_{xy}$ orbitals of a single As atom. The $p_y/p_x$ bands around the Fermi level are found to carry the opposite spin. The splitting between the spin-up and spin-down pockets of the same orbital is used to estimate $J_{pd}$. (b) Schematics of the $p-p$ hopping and $p-d$ exchange.  (c) Downfolding the As-$p_{xy}$ bands to the minimal model. upper: the DFT bands; middle: the As-$p_{xy}$ bands on the basis of Maximally localized Wannier functions; bottom: the $p_x$ bands from the minimal model in Eq.(\ref{Hp}).}
\label{fig-5}
\end{figure*}

\section{Itinerant holes and their coupling to local spins}
The previous section focuses on the localized Fe $d$-electrons. We now turn to the itinerant As $p$-electrons lying right below the Fermi level. Due to the small charge-transfer gap, charge fluctuation between the ground-state $d^5p^6$ configuration and the excited $d^6p^5$ configuration is possible at ambient temperature. The activated mobile holes associated with the itinerant $p$-bands are considered to dominate the charge transport.

The hole valley centered at the M-point arises from the \emph{in-plane} As $p$-orbitals, as shown in Fig. \ref{fig-5a} by the orbital- and spin-resolved band structure. The principle axes of the in-plane $p$-orbitals are chosen along the As-As bonding directions, which rotate by 45 degrees with respect to the $\textbf{a}_1$-$\textbf{a}_2$ axes. The interesting point is that under such a projection the $p_x$ and $p_y$ electrons are nearly decoupled around the band edge. Additionally, they carry the opposite spin [Fig.\ref{fig-5a}]. In other words, the hole carriers feature unique orbital-dependent spin polarization, as illustrated by a schematic plot in Fig. \ref{fig-5b}.

To reveal the physical origin, we specify As $p_{xy}$ orbitals as the starting point to construct the corresponding maximally localized Wannier functions out of the valence-band Bloch wave functions. We employ the ``disentanglement'' procedure introduced in Ref.\cite{disentangle} to separate out the Fe $d$-, Cu $d$- and As $p_z$-dominated bands. The resulted energy bands spanned by the As $p_{xy}$-like Wannier functions are plotted in Fig \ref{fig-5c}, which nicely reproduces the overall dispersion of the valence bands. Note that this optimal subspace consists of (2 orbitals/As )$\times$ (2 As/layers) $\times$ 4 layers = 16 bands in total. Those localized $d$-bands [c.f. Fig.\ref{fig-1b}] are automatically projected out.
A minimal model can be written by neglecting the hopping terms between the $p_x$ and $p_y$ orbitals and coupling between the different As layers:
\begin{eqnarray}\label{Hp}
H_{p_x}=\mu \sum_i c_{x,i}^\dagger c_{x,i}+t_\sigma \sum_i c_{x,i}^\dagger c_{x,i+\frac{\textbf{a}_1+\textbf{a}_2}{2}} \\ \nonumber
+t_\pi \sum_i c_{x,i}^\dagger c_{x,i+\frac{\textbf{a}_1-\textbf{a}_2}{2}}+H.c. ,
\end{eqnarray}
where $\mu$ is the $p_{xy}$-orbital chemical potential, which rigidly shifts the band energy and determines the top of the hole bands from the Fermi level. $H_{p_y}$ can be obtained by simply reversing $t_\sigma$ and $t_\pi$. Fig. \ref{fig-5c} (bottom panel) shows the valence band dispersion from the minimal model with $t_\sigma$=-0.9 eV and $t_\pi$=0.3 eV. Notwithstanding the simplicity, the hole valley at the M point and the total $p$-band width (c.f. $W_p$ in Table \ref{table-1}) are correctly described.

From Eq.(\ref{Hp}), the difference between the $p_x$ and $p_y$ electrons becomes clear. Due to the orbital anisotropy, $p_x$ and $p_y$ electrons form strong $\sigma$ bonds along the $[1\bar{1}0]$ and $[110]$ directions, respectively. Note that these two perpendicular directions cut different Fe atoms in the AFM spin chain [Fig.\ref{fig-5b}]. A crosscheck reveals that the spin direction of the  $p_{xy}$ holes are parallel to that of the intersecting Fe. Thus, the orbital-dependent spin polarization of holes can be explained by the directional Hund's coupling to different sublattices of the AFM chain, which can be described by:
\begin{eqnarray}
H_{pd}=-J_{pd}\sum_{\gamma,\langle i,j \rangle_\gamma} \textbf{s}_{\gamma i} \cdot \textbf{S}_j ,
\end{eqnarray}
where $\gamma=x,y$, $\textbf{s}_{\gamma i}=\sum_{\alpha\beta}c_{\gamma i\alpha}^+\vec{\sigma}_{\alpha\beta}c_{\gamma i\beta}$ and $\langle ij \rangle_\gamma$ denotes the nearest neighbor sites along the $\gamma$-direction. The Hund's coupling $J_{pd}$ arises from the overlap between the As $p_{xy}$ orbitals and the Fe $d$-orbitals. We roughly estimate $J_{pd}\sim0.5$eV by referring to the energy splitting between the spin majority/minority valleys [See the horizontal lines marked in Fig. \ref{fig-5a}].

It is reasonable to speculate that the spin-polarized charge current exists along the $[1\bar{1}0]$ or $[110]$ direction. The subtlety here is that the top and bottom As layers of a As-Fe-As sandwich have exactly the opposite orbital polarization, as dictated by the inversion symmetry. To obtain a net spin current, one needs to break this symmetry, e.g. by applying a perpendicular electric field. Breaking the degeneracy between the $p_x$ and $p_y$ orbitals, e.g. by applying an uniaxial strain along the $[110]$ direction, should enhance the degree of spin polarization. Furthermore, due to the spin-hole coupling, magnetoresistance may also be observed.

\section{Discussion and Conclusion}
By combining all the results above, the complete low-energy model of NaFe$_{0.5}$Cu$_{0.5}$As can be written as:
\begin{eqnarray} \label{H}
H=H_d+H_p+H_{pd}
\end{eqnarray}
A fundamental difference between NaFe$_{0.5}$Cu$_{0.5}$As and cuprates is the Hund's coupling between the hole and the local spin. Recall that when a hole is doped into high-Tc cuprate superconductors, it goes predominantly into a 2p orbital of an oxygen site. Together with the hole on a Cu site, it forms a singlet state commonly named after Zhang and Rice,\cite{ZhangRice1988} which has been considered as a starting point to discuss the microscopic origin of the normal state and superconducting properties of cuprates. Here, in NaFe$_{0.5}$Cu$_{0.5}$As, due to the ferromagnetic couplng, the holes on the p-orbital of As do not bind with the Fe spin into singlets. This difference may be due to the nonplanar Fe-As bonding geometry and the large spatial extension of the As 4$p$-orbitals. In some sense, NaFe$_{0.5}$Cu$_{0.5}$As appears more like a narrow-gap magnetic semiconductor. The thermally activated mobile holes associated with the itinerant As $p$-bands carry charge current, and their spins can be polarized by the underlying magnetic ions. It will be interesting to see if this parent compound NaFe$_{0.5}$Cu$_{0.5}$As  can be hole doped. By reducing the hole excitation energy to zero, the system becomes a typical ``Hund's metal'', which has been extensively studied in the context of iron-based superconductors. \cite{{yin2010magnetism},{yin2011kinetic}} A similar two-fluid model as Eq.(\ref{H}) is considered to spawn an intricate interplay of nematicity, spin-density wave and superconductivity.\cite{you2014twofuild}

Our discussion so far does not take into account the charge fluctuation of Fe. We assume that when the temperature is not high, the small concentration of thermally excited electrons does not destroy the AFM order of the Fe chains. A rigorous study is, however, beyond the capability of the DFT+U formalism. This problem is equivalent to electron-doping the half-filled quasi-1D Fe chains (the As $p$-bands become irrelevant). We refer to a related density matrix renormalization group calculation, which reveals exotic magnetic order within the orbital-selective Mott regime. \cite{PRB2016DMRG} This scenario in Cu-substituted iron-based superconductors is discussed in Ref. \cite{NJP2016OSMP}.

\section*{Acknowledgements}
We thank Yu Song, Pengcheng Dai, Yayu Wang, Hong Yao and Zhengyu Weng for helpful discussion. Z.L., S.Z. and Y.H. acknowledge Tsinghua University Initiative Scientific Research Program. S.Z. is supported by the National Postdoctoral Program for Innovative Talents (BX201600091) and the Funding from China Postdoctoral Science Foundation (2017M610859). J. M. and F. L. acknowledge the support from US-DOE (Grant No. DE-FG02-04ER46148). The primary calculations were performed on the Tianhe-II supercomputer provided by The National Supercomputer Center in Guangzhou, China. PARATERA is also sincerely acknowledged for the continuous technical support in high-performnace computation .

%

\bibliography{ref}

\end{document}